\documentstyle[12pt]{article}
\def \medio  {\baselineskip= 1.5 \normalbaselineskip}

\newcommand{\titul}[1] {\begin{center}{\large\bf #1 } \end{center}\vskip 1.cm}
\newcommand{\autor}[1] {\begin {center} {\large \lineskip .5em #1 }
                        \end   {center} }
\newcommand{\lugar}[1] {\begin{center} {\it #1} \end{center}}
\newcommand{\abstr}[1] {{\begin{center} \vskip .5cm {\bf Abstract
                        \vspace{0pt}} \end{center}}\begin{quote} #1
                        \end{quote}}

\begin{document}
\begin{titlepage}
\begin{flushright} {\bf US-FT/10-00 } \end{flushright}

\vskip 3.cm
\titul{
Nuclear Effects on the UHE Neutrino-Nucleon\\
Deep Inelastic Scattering Cross Section
}
\autor{J.A. Castro Pena, G. Parente and E. Zas}
\lugar{Departamento de F\'\i sica de Part\'\i culas\\
Universidade de Santiago de Compostela\\
15706 Santiago de Compostela, Spain}
\abstr{
\medio

Using a recent parametrization of nuclear effects in
parton distribution functions we calculate the
neutrino-nucleon cross section at
energies relevant for ultra high energy neutrino telescopes.
The modification of the cross section in comparison with the
calculation using parton densities in free nucleons is of the
order of few per cent for the parameter range of interest
in neutrino telescopes
(A=10 and E=10$^6$ GeV) and it reaches 20 $\%$ at the highest energies
(E=10$^{12}$ GeV) and for the largest nuclear size (A=190) considered.
}
\end{titlepage}
\newpage

\pagestyle{plain}
\medio
\section{Introduction} \indent

The high energy neutrino cross section is a crucial ingredient in the
calculation of the event rate in high energy neutrino telescopes. 
The detectors presently in design or construction stages \cite{AMANDA}
look for \v Cerenkov light from the neutrino-induced muon in charged 
current interactions and take advantage of both the long muon range 
and the rise of the neutrino cross to meet the neutrino detection challenge. 
The background of atmospheric muons is rejected searching for neutrinos
that have travelled through the earth. For energies above a few hundred 
TeV the interaction length for neutrinos becomes comparable to 
the Earth diameter. As a result the event rate is a convolution of  
the differential spectrum, the cross section and the 
exponential attenuation of the neutrino flux. 
Uncertainties in the cross section get amplified into the energy 
spectrum and the angular distribution of the event rate and it is 
important that they are kept under control. 

The DIS cross-sections are obtained from $x$ and $Q^2$ dependent
structure functions which are calculated in perturbative QCD.
In practice, structure functions are obtained from
$x$ and $Q^2$ dependent parton distribution functions
which follows DGLAP evolution equations \cite{DGLAP}.
The DGLAP formalism effectively resum logarithmic terms in
$Q^2$ which appear in the perturbative series.

The calculation of the neutrino-nucleon cross section 
involves an integration over $x$. 
As the energy increases, the integral becomes dominated by
the interaction with low $x$ quarks and gluons 
while the average $Q^2$ rises to values of order 
the electroweak boson mass squared (for a detailed discussion see
for example Ref. \cite{ymedio}).
Then, in the process of the total cross section 
calculation, the parton densities extracted from present data must
be extrapolated using the theoretical formulas to the region in the 
$x, Q^2$ plane not supported by experiments. 
 
The uncertainty due to the extrapolation 
to high $Q^2$ is not expected to be large because the 
$x$ shape of the parton distribution functions at low 
$Q^2$ is narrowly constrained by HERA data while the further
evolution to high $Q^2$ values seems to be well defined in DGLAP. 
At the highest energies, uncertainties between 
$\pm 20\%$ \cite{Gluck} and a factor 
$2^{\pm 1}$ \cite{Reno} are typically reported. 

However, DGLAP is expected to break down at low $x$
because of potentially large logarithmic terms in $x$ which also 
appear in the perturbative series. 
These dangerous logarithms are partially considered under certain
approximations in DGLAP.
A different approach is provided by the BFKL formalism
\cite{BFKL} which deals with the $Q^2$ evolution
of unintegrated in transverse momentum parton densities.
The BFKL predictions are nowadays under discussion:
the leading order (LO) result does not agree with data and
the next-to-leading order (NLO) corrections are found to be suspiciously
large to be trusted, although it seems that the adequate choosing of
the renormalization scale
\cite{KIM} can solve this problem.
 
We discuss in this letter a different source of uncertainty
in the calculation of the high energy neutrino-nucleon DIS cross section,
which comes from the fact that parton densities widely used were
extracted under the assumption that partons belong to isolated nucleons
when, on the contrary, the nucleons are usually bound in larger
nuclei at the interaction sites.
In this work we perform a more realistic calculation taking into account
these effects by using the standard parton densities in free nucleons
with the modifications due to nuclear effects as given by the EKS
parameterizations \cite{EKS}.

\section{Nuclear effects}
\indent

It is experimentally well known that the structure function $F_2$ 
in deep inelastic lepton-nucleus scattering for large atomic mass, $A$,
is different from that measured for hydrogen or deuterium targets
(see for example \cite{NMC}).
If one maintains the partonic view
of the nucleon, this result may indicate that
parton distributions of bound nucleons are different from
those of free nucleons.

Parton distributions of the nucleon in a nucleus containing
$A$ nucleons have been obtained in Ref. \cite{EKS} using
QCD evolution at leading twist and leading order of perturbation
theory together with the experimental ratios $F_2^A/F_2^D$ 
measured by EMC and NMC at CERN \cite{NMC,data}.

With the quark ratios as defined in Ref. \cite{EKS}: 
\begin{eqnarray}
R_V^A(x,Q^2) \equiv 
\frac{u_V^A(x,Q^2)+d_V^A(x,Q^2)}{u_V(x,Q^2)+d_V(x,Q^2)}
\;\;\;\;\;\; \mbox{and}
\\ \nonumber \\
R_S^A(x,Q^2) \equiv \frac{{\bar u}^A(x,Q^2) + {\bar d}^A(x,Q^2) +
{\bar s}^A(x,Q^2)}{{\bar u}(x,Q^2) + {\bar d}(x,Q^2) + {\bar s}(x,Q^2)} \, ,
\\ \nonumber
\end{eqnarray}
we have generated the correction due to nuclear effects
to the parton distribution functions in free nucleons.
The ratios of valence and sea quarks given
by Eq. (1) and (2) are plotted in Fig. 1. From low to high $x$
one can easily identify in Fig. 1 the regions of
shadowing ($R<1$), anti-shadowing ($R>1$), EMC ($R<1$) and Fermi
motion ($R>>1$) which are experimentally observed in the ratio 
$F_2^A/F_2^D$. We show below that the shadowing region (small $x$) 
gives the main correction to the neutrino-nucleon DIS cross section
at high energy.
The smooth $Q^2$ dependence of the ratio, which was difficult to
observe experimentally, is also apparent in Fig. 1.

We have calculated the neutrino-nucleon DIS
cross-sections following Ref. \cite{ymedio}\footnote{For brevity,
in this letter we do not present the explicit expressions for the
cross sections calculation}
but
considering valence and sea parton distributions of the nucleus as
corrected by Eqs. (1) and (2).
In the calculation we use parton densities from the
group MRST \cite{MRST98} at the LO approximation in $\alpha_s(Q^2)$.
At the LO approximation we are
consistent with the EKS parameterization of nuclear effects \cite{EKS}
which has been extracted with the help of LO evolution equations
equations. However, the effect of higher order corrections should be
small \cite{ICRCDURBAN} and it would not modify the conclusions of the
present work.
In addition, in the cross section calculation we have also neglected the
contribution from
the longitudinal structure function, which is proportional
to $\alpha_s(Q^2)$, and terms suppressed by 
the ratio of the nucleon mass to $Q^2$.
 
Let us comment some technical details of the calculation:  
In the integration we take the minimum value of Q$^2$=4 GeV$^2$
although the result is not sensitive to variations around this value.
We have also extrapolated the $x$ dependence of the quark
distribution functions
beyond the limits of applicability given by the authors 
using the $x$ behavior at the lowest $x$ value of the parametrization 
For the MRST98 set, we assume the Regge type shape $\sim x^{-\lambda}$
below $x = 10^{-5}$ with $\lambda$ the slope obtained at $x = 10^{-5}$.
This simple phenomenological approach agrees with the
extrapolations based on perturbative QCD (for example
the double-logarithmic-approximation) which explain HERA data.
For the EKS parameterizations, we take below $x = 10^{-6}$ the
constant behavior shown in Fig. 1 at low $x$.

In this work we are also interested in the modification due to
nuclear effects of the average value of the inelasticity $y$,
which is the fraction of
the neutrino energy which flows to the hadronic part of the interaction 
in the laboratory frame.

This parameter fixes the relative rates of 
the two main types of detections in neutrino telescopes,
using muons in charged current interactions and using the showers.
It is also responsible for the 
relative sizes of the electromagnetic and hadronic showers induced 
in a charged current electron neutrino interaction.
The knowledge of $y$ is also necessary for extracting the neutrino 
energy in high energy neutrino telescopes from the detected 
muon or hadronic shower. 
On the other hand it has been found that $\langle y \rangle$
is directly related to the $x$ slope of the parton distribution functions
at small $x$ and $Q^2$ around $M_W^2$ \cite{ymedio}.
If $\langle y \rangle$ could be extracted from events observed in
high energy neutrino telescopes \cite{Jaime}, it should provide a
unique oportunity to study the low $x$ physics at high $Q^2$, well beyond
the reach of present and near future accelerators.

It is very interesting to study the stability of both, the total
cross section and $\langle y \rangle$, to the consideration of nuclear
effects in parton distributions.
The average value of $y$ is obtained by integrating the 
differential cross-section:
$
\langle y \rangle = 1/\sigma \int_0^1 dy \, y \, d\sigma/dy
$.
The results for the total cross-section and the average $y$
are compared to the non-corrected ones in Figures 2 and 3.
From Fig. 2 one sees that the departure from the total cross-section
for free nucleons increases with energy and $A$ at the highest energies.
For example, for $E_{\nu}=10^{10}$ GeV, CC interaction and
$A=60$, the modified
parton distribution functions result in a 10 \% reduction of the
total cross section. The regions of EMC effect, anti-shadowing
and shadowing are apparent in Fig. 2.

The modification of $\langle y \rangle$ due to nuclear effects
is presented in Fig. 3 for the case
of neutrino-nucleon CC interaction. 
The  correction to the results with partons in a free nucleon 
is less than 3 \% for the largest nucleus
and neutrino energies around 10$^7$ GeV.

In conclusion, we have calculated the UHE neutrino-nucleon DIS cross
section with parton distributions in bound nucleons to take into
account nuclear effects. In comparison
with the calculation using partons in free nucleons we found
that the correction, which is mainly due to shadowing
at small $x$, is of the
order of few per cent for the parameter range of interest
in neutrino telescopes
(A=10 and E=10$^6$ GeV) and it reaches 20 $\%$ at the highest energies
(E=10$^{12}$ GeV) and for the largest nuclear size (A=190) considered.

\vspace{1cm}
\hspace{1cm} \Large{} {\bf Acknowledgements}    \vspace{0.5cm}

\normalsize{}
This work was supported by Xunta de Galicia
under grant PGIDT00-PXI20615PR and CICYT grant AEN99-0589-C02-02.


\vspace{2.cm}

\hspace{1cm} {\Large{\bf Figure captions}}    \vspace{0.5cm}

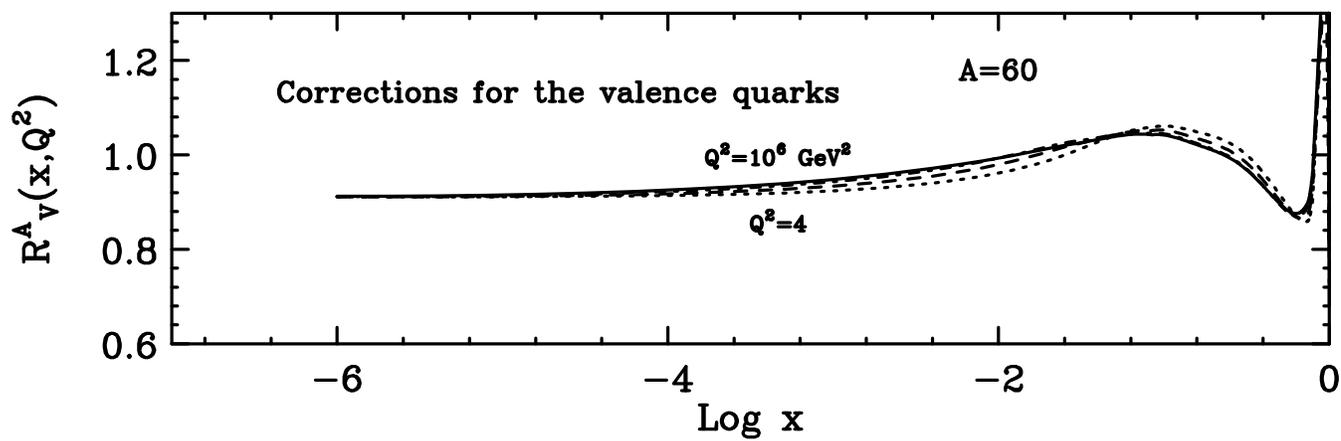
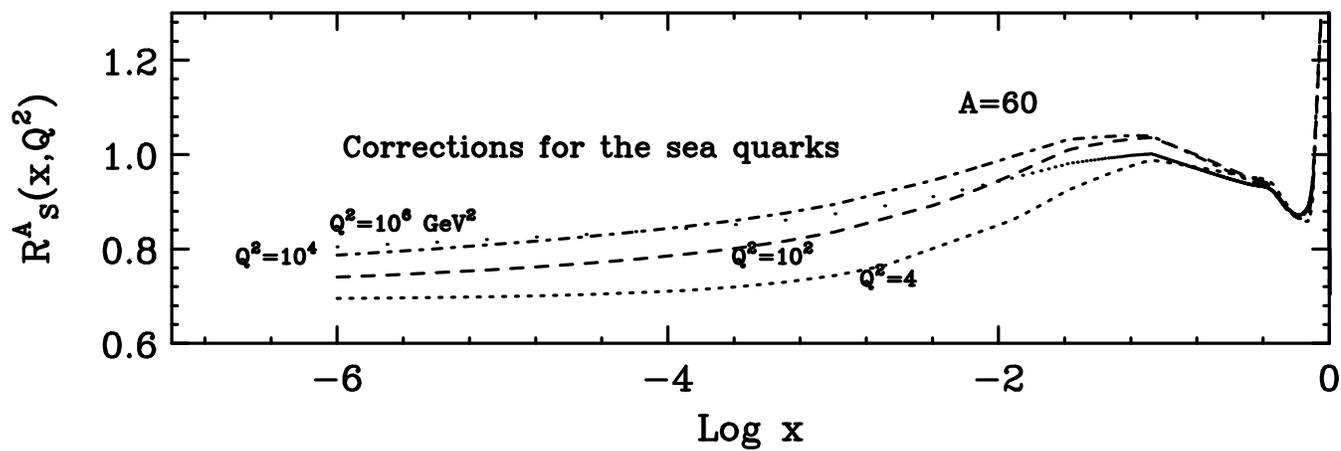
{ \bf Figure 1.} The ratio of nuclear
parton (valence and sea quarks) distribution functions over free parton
distributions as a function of the $x$-Bjorken variable for different
values of $Q^2$

\vspace{0.5cm}

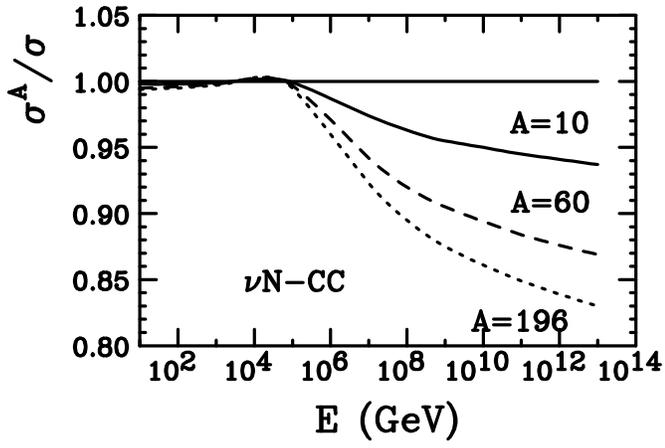
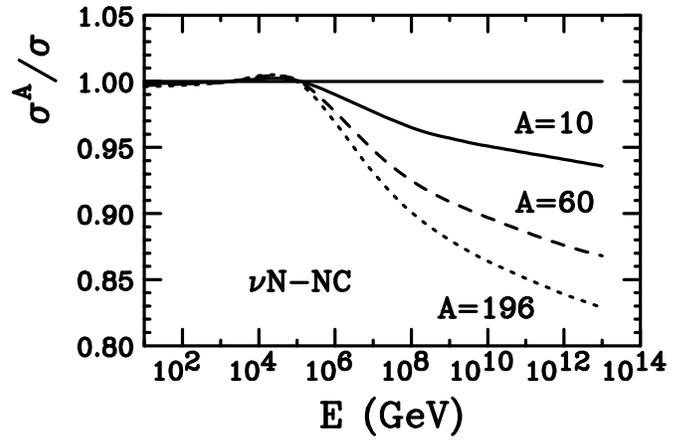
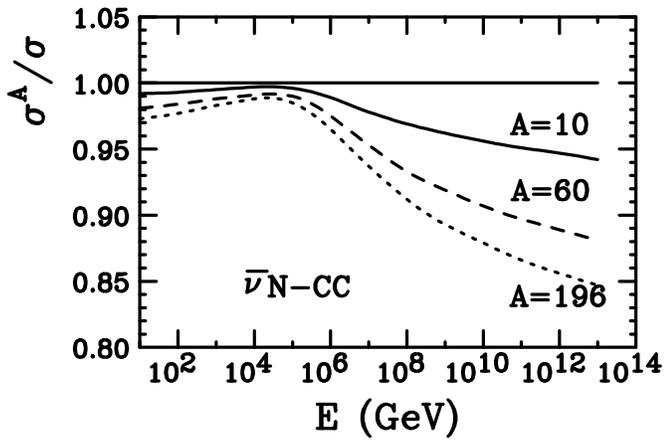
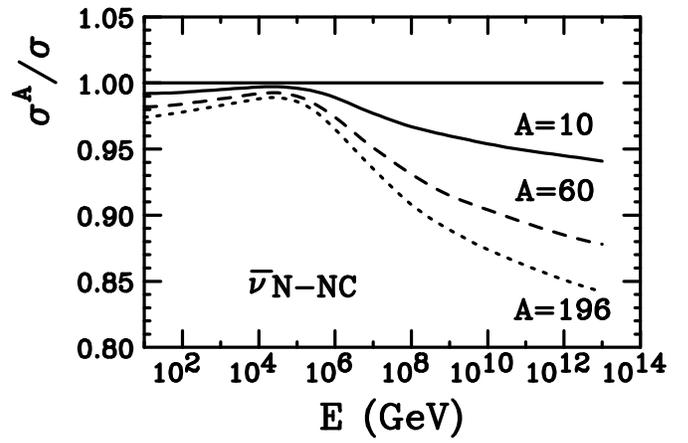
{\bf Figure 2.} The ratio of nuclear effects corrected over non corrected
total neutrino-nucleon cross sections
as a function of neutrino energy in the lab frame for different
nuclei A.

\vspace{0.5cm}

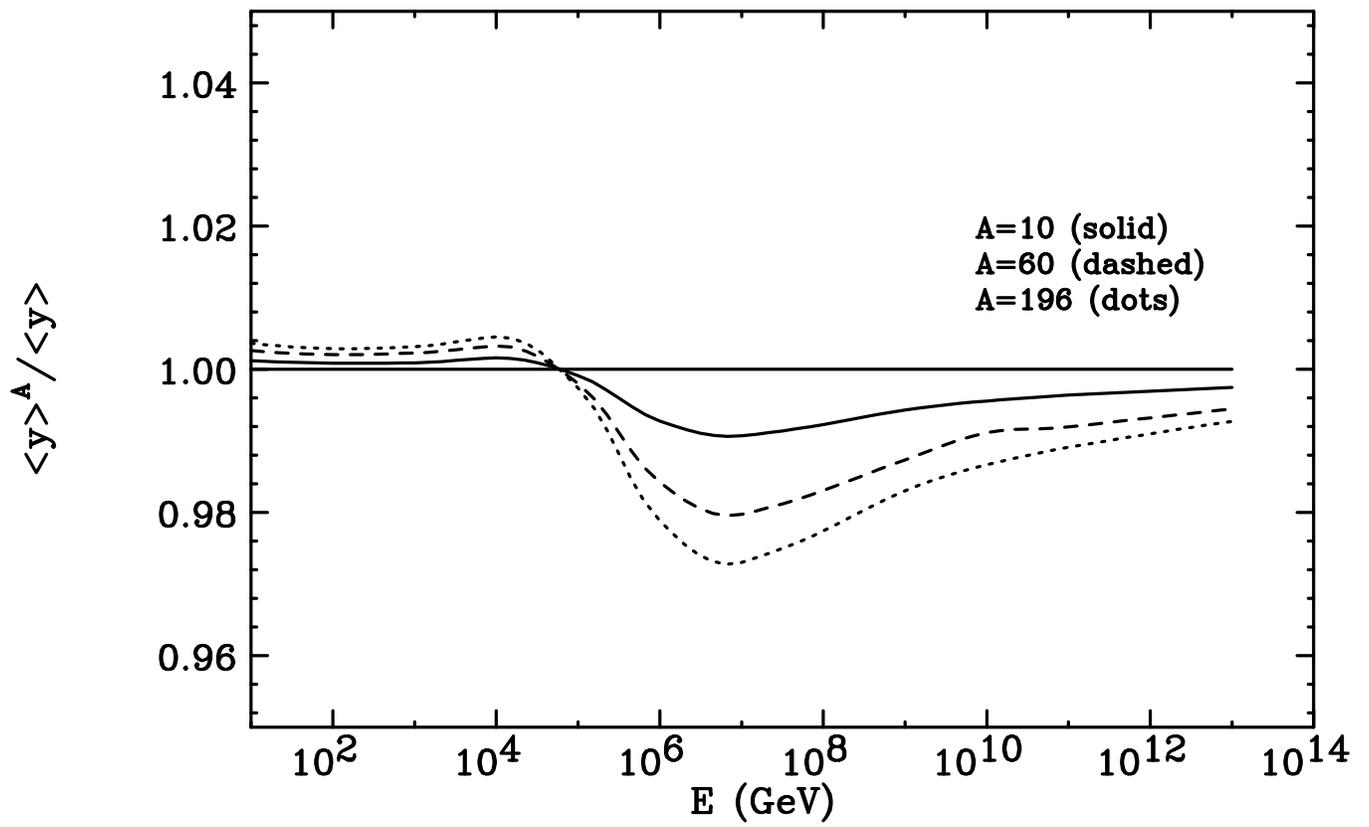
{ \bf Figure 3.} The ratio of nuclear effects corrected over non corrected
average inelasticity in CC $\nu N$ interaction for different nuclei A
as a function of neutrino energy in the lab frame.

\end{document}